\RequirePackage{fix-cm}
\documentclass{svjour3}                     
\smartqed  

\usepackage{graphicx}
\usepackage{amssymb}
\usepackage{graphicx}
\usepackage[space]{grffile}
\usepackage{latexsym}
\usepackage{textcomp}
\usepackage{longtable}
\usepackage{multirow,booktabs}
\usepackage[]{booktabs}
\usepackage[]{graphicx}
\usepackage[]{color}
\usepackage[]{array}
\usepackage[]{hyperref}
\usepackage[utf8]{inputenc}
\usepackage[english]{babel}
\usepackage{csquotes}
\hypersetup{
    colorlinks,
    citecolor=blue,
    filecolor=blue,
    linkcolor=blue,
    urlcolor=blue
}
\definecolor{banana}{RGB}{255, 225, 53}
\newcounter{EKXCommentsCounter}

\newcommand{\myquote}[1]
{
  \begin{quotation}
  #1
  \end{quotation}
}

\newcolumntype{L}[1]{>{\raggedright\let\newline\\\arraybackslash\hspace{0pt}}m{#1}}
\newcolumntype{C}[1]{>{\centering\let\newline\\\arraybackslash\hspace{0pt}}m{#1}}

\begin{document}

\title{Software Engineering in Start-up Companies \textendash Analysis of 88 Experience reports}




\author{Eriks~Klotins \and Michael~Unterkalmsteiner \and Tony~Gorschek}



\institute{E. Klotins, M. Unterkalmsteiner, T. Gorschek \at
              Software Engineering Research Lab \\
              Blekinge Institute of Technology \\
              \email{eriks.klotins@bth.se}           
}

\date{Received: date / Accepted: date}

\maketitle

\begin{abstract}

\textit{Context:} Start-up companies have become an important supplier of 
innovation and software-intensive products. The flexibility and reactiveness of 
start-ups enables fast development and launch of innovative products. However, a majority of 
software start-up companies fail before achieving any success. Among other factors, 
poor software engineering could be a significant 
contributor to the challenges experienced by start-ups. However, the 
state-of-practice of software engineering in start-ups, as well as the 
utilisation of state-of-the-art is largely an unexplored area. 
\\\textit{Objective:} In this study we investigate how software engineering is applied in start-up context with a focus to identify key knowledge areas and opportunities for further research. 
\\\textit{Method:} We perform a multi-vocal exploratory study of 88 start-up 
experience reports. We develop a custom taxonomy to categorize the reported 
software engineering practices and their interrelation with business aspects, 
and apply qualitative data analysis to explore influences and 
dependencies between the knowledge areas.
\\\textit{Results:} 
We identify the most frequently reported software engineering (requirements 
engineering, software design and quality) and business aspect (vision and 
strategy development) knowledge areas, and illustrate their relationships. We 
also present a summary of how relevant software engineering knowledge areas are 
implemented in start-ups and identify potentially useful practices for adoption 
in start-ups.  
\\\textit{Conclusions:}  The results enable a more focused research on 
engineering practices in start-ups. We conclude that most engineering 
challenges in start-ups stem from inadequacies in requirements engineering. 
Many promising practices to address specific engineering challenges exists, 
however more research on adaptation of established practices, and validation 
of new start-up specific practices is needed.

\end{abstract}

%
%
%
%

\section{Introduction}
\label{sec:introduction}

Software start-ups are important suppliers of innovation and innovative software 
products~\cite{Baskerville2003}, providing products and 
services that are a significant part to the 
economy~\cite{StartupCompassInc.2015}. This potential is strengthened further 
as the 
use of cutting-edge technologies enable start-ups to develop, launch and evolve 
software products fast and with very few resources~\cite{Baskerville2003}.

A challenge is that most start-up companies collapse before any significant 
achievements are realized~\cite{Tovstiga2012}. This is explained by market 
conditions, lack of commitment, financial issues or, simply put, a bad product 
idea. However, product engineering activities takes substantial resources from start-ups, especially in the early stages~\cite{Crowne2002,Giardino2015}.
Inadequacies in used engineering practices could lead to under or over-engineering the product, wasted resources, and missed market opportunities. 

One of the main qualities of start-ups is their ability to quickly take advantage of new business, market and technology opportunities~\cite{Giardino2014,bajwa2017failures}. Decisions, such as what features to build, how and when, belongs to the realm of engineering and have a huge impact on how the start-up responds to new opportunities. For example, certain decisions may hamper flexibility of the product, thus reducing the speed of adapting the product for entering new markets. Taking one sub-optimal decision may have only a small effect on start-up's prospects, however the compound effect of the decisions determines whether the start-up is able to remain on the edge of innovation or is struggling to keep its product running. This is a source of risk and opportunity with potential effects to all aspects of the company.

Yau et al.~\cite{Yau2013} argue that practices adapted from established 
companies attempt to solve problems that are not present in start-ups, while ignoring start-up 
specific challenges, such as time-to-market as the primary goal, and 
accumulating technical debt~\cite{Unterkalmsteiner}. Even though 
similar challenges can be 
present in established
organizations too and addressed by state-of-the-art practices, it is the 
combination of multiple challenges that makes engineering in start-ups 
difficult. There is a gap in understanding how these start-up specific challenges influence the 
engineering process and what engineering practices are suitable for such 
context~\cite{Klotins2015,Paternoster2014a}.

This lack of understanding results in that there are very few, if 
any, start-up context relevant software engineering processes/methods/models/frameworks 
(called practices from now on). At the same time, a substantial amount of money 
is invested in start-up companies.
In the first three quarters of 2015 alone, 
start-up companies received investments 
of 429 billion USD in the US and Europe~\cite{PitchBookData2015a,PitchBookData2015}. 
With an optimistic start-up failure rate of 75\% this constitutes 322 billion USD of capital potentially wasted on building unsuccessful 
products~\cite{StartupCompassInc.2015,PitchBookData2015a,PitchBookData2015}. To what 
extent inadequacies in software engineering practices are responsible or linked to success 
rate is very hard to judge. However, even if the effect of improved engineering practices 
only would result in a few percent change in success rate, it would yield significant 
impact and capital return. Thus, the focus of our study is to explore specifically software 
engineering in start-ups and pinpoint specific areas for further research that are likely to
benefit start-up practitioners.

Researchers have recognized the importance of software engineering in 
start-ups. Bosch et 
al.~\cite{Bosch2013} and Deakins et al.~\cite{Deakins2005} propose adaptations of 
iterative and incremental development methods to address engineering challenges in 
start-up companies. However, this work is preliminary and has not been validated yet
in practice~\cite{Klotins2015,Paternoster2014a}. Giardino et al.~\cite{Unterkalmsteiner} 
report on an interview study aiming to understand how start-ups 
select their product development strategy and how start-ups consider product quality 
attributes. Giardino et al.~\cite{Giardino2015} also investigated the key challenges in 
software start-ups and report that technology uncertainty is the key challenge in 
software start-ups. However, none of these studies provide a comprehensive answer of what 
engineering practices are relevant in start-ups.

In this paper we use empirical data from 88 start-up experinece reports to provide the first insight into what engineering knowledge areas are relevant in software start-ups. Our aim is to explore what engineering practices the start-ups report as relevant, how these practices are applied and what results they yield. To analyze the reports we use qualitative data analysis methods~\cite{Seaman1999,Garousi2016}

To cater for the fact that the experience reports cover also business and 
marketing aspects, which are tightly intertwined with software engineering 
aspects, we use a software and business practices taxonomy to support the 
analysis of the reports. Even though we acknowledge importance of good business, market and other practices, the focus of this paper is strictly on software engineering practices.

The main contribution of this paper is the identification and description of the state-of-practice in software start-up companies, pinpointing to several relevant software engineering areas that need further research. Moreover, we present related work to each relevant software engineering knowledge area, illustrating potentially useful engineering practices for start-ups.

The remainder of this paper is structured as follows: 
Section~\ref{sec_related_work} 
presents background and related work and Section~\ref{sec_rm} introduces the 
research methodology. The results are presented in Section~\ref{sec_results}, 
analysed and discussed in Section~\ref{sec_analysis}. 
Section~\ref{sec_conclusions} concludes the paper.

\section{Background and related work} \label{sec_related_work}

\subsection{Software start-ups}

As early as 1994, Carmel~\cite{Carmel1994c} recognizes small software companies 
being exceptionally successful at innovation and delivery of new products. 
Termed software start-ups, these companies share many features with small and 
medium enterprises such as market pressure, youth and immaturity, and limited 
resources~\cite{Sutton2000}. However, start-up companies are different due 
their goals and challenges. Contrary to established companies who aim to shape 
their products to address a known market need, start-up companies attempt to 
identify an unmet market need and to invent a product satisfying this 
need~\cite{StartupCompassInc.2015}.

The engineering context in start-ups is characterized by uncertainty, lack of 
resources, rapid evolution and an immature team among other 
factors~\cite{Sutton2000}. However, the start-up context also provides 
flexibility to adapt new engineering practices and reactiveness to keep up 
with emerging technologies and markets~\cite{Giardino2014}.

Product related issues are reported as the key challenge in start-up 
companies~\cite{Giardino2015}. However, recent literature mapping studies identify a lack 
of research in the area from a software engineering 
perspective~\cite{Klotins2015,Giardino2014}. Moreover, most publications to date are 
experience reports lacking in-depth analysis and rigorous research 
methods on empirical data~\cite{Klotins2015}.

Despite attempts to explore the start-up phenomenon, only a few studies 
specifically focus on understanding how software engineering is done in 
start-up companies. Yau et al.~\cite{Yau2013} and Sutton~\cite{Sutton2000} recognized 
that engineering practices aimed at established 
companies are not suitable for start-ups. As a result, various models for 
software development in start-up context were proposed~\cite{Deakins2005,Bosch2013,Zettel2001}, however there is very little evidence of application and validation of these models~\cite{Klotins2015}.

An agenda to specify important research topics regarding software engineering 
in start-ups has been created by the Software Start-up Research 
Community~\cite{Abrahamsson2016}. Among other topics, engineering practices in 
start-ups is identified as an important research area.

Rafiq et al.~\cite{Rafiq} and Melegati et al.~\cite{Melegati} have studied 
requirements engineering practices in start-ups and provide an insight into how 
product ideas evolve and what practices are used to connect founders' vision 
with customer needs.  

Crowne~\cite{Crowne2002} proposes a start-up life-cycle model and identifies 
goals and the key challenges at each of the life-cycle phases. According to the 
model, in the first phase a company develops an early version of the product. The goal of 
the second phase is to improve quality of the product until it can be 
commissioned with little effort. In the third phase, the company grows and 
conquers the market. In the fourth phase the company matures into an established 
organization. However, there is no detailing on the use or adaption of engineering 
practices, or their evolution over time and phases.

Giardino et al. present a behavioral model aimed at
explaining start-up failures. They argue that a start-up must first explore the 
problem domain and then validate a proposed solution, however mismatching 
validation activities with exploration activities could lead to a 
failure~\cite{Giardino2014a}. This is directly related to engineering practices such as 
overall requirements engineering or scoping.

Olsson et al.~\cite{Olsson2015} identify challenges of using customer 
feedback in large software-intensive product engineering. A case study reveals that 
to compensate for inadequate utilization of user feedback, companies invent 
requirements and steer product direction by ``gut feeling''. 
Invented requirements lead to a large amount of unused and incorrectly 
implemented features contributing to product failure~\cite{Olsson2015}.

Software engineering in start-ups shares many similarities with companies using 
agile 
development practices, such as iterative development, empowered small team, and ongoing 
planning~\cite{Ramesh2007,Chow2008a}. However, customer involvement, which is one of the 
key agile principles, is difficult to establish as start-up companies lack a 
distinct set of customers. Hence, start-ups operate in a market-driven 
environment~\cite{Dahlstedt}. In a market-driven environment, requirements are 
often invented and validated through frequent product 
releases~\cite{Dahlstedt,Alves2006}.

There has been a substantial work to identify general operating practices for 
start-ups, such as The Lean Start-up \cite{Ries2011} and Customer Development 
Model \cite{Blank2013b}. However, there exist very few peer-reviewed studies in 
software engineering fora, with May \cite{May2012} being the exception,  
reporting on the application of said practices. Therefore, it remains to be 
explored to what extent these or any other practices are adopted by start-ups.

\subsection{Scope of software engineering in start-ups}

Sutton~\cite{Sutton2000} argues that start-up companies are sensitive to many 
influencing factors, such as customers, partners, changes in technologies and 
markets. The developed software is often a component of 
another product and which is an essential part of the start-up company 
itself~\cite{Baskerville2003}. Therefore, software engineering in start-ups must be 
studied jointly with its dependencies and influences to other areas of a start-up~\cite{Broy2006}.

The interrelation of software engineering, product development and other areas 
in an organization is recognized by ISO/IEC 42010:2011 which provides a 
definition of software-intensive system as: ``any system where software 
contributes essential influences to the design, construction, deployment, and 
evolution of the system as a whole to encompass individual applications, 
systems in the traditional sense, subsystems, systems of systems, product 
lines, product families, whole enterprises, and other aggregations of 
interest''~\cite{IEEE2011}.

To fully understand the importance of software engineering in start-ups we 
widen our field of view and capture other concepts that influence or depend 
on software engineering. To define the scope for this study we use a taxonomy, 
further elaborated in Section~\ref{sec_swbp}.

\subsection{Software engineering and business practice taxonomy} \label{sec_swbp}

A taxonomy is useful for categorization and mapping of knowledge, 
facilitating identification of gaps and boundaries of a 
phenomenon~\cite{Smite2014}. Seaman~\cite{Seaman1999} suggests to use of 
preformed codes, e.g. a taxonomy, to 
facilitate coding in qualitative studies.
Hence, to support the identification of software 
engineering and other relevant practices we developed a taxonomy listing 
knowledge areas and practices for start-ups - the SoftWare and Business Process, hereinafter the SWBP, taxonomy. The taxonomy consists of 
software engineering knowledge areas as defined by SWEBOK, and several business aspects oriented knowledge areas inspired by Osterwalder~\cite{Osterwalder2005} and Zachman~\cite{Zachman}.

Even though SWEBOK is not specifically designed for software 
start-up companies and lacks emerging areas of software engineering, such as 
value-based software engineering~\cite{Boehm2003,Azar2007} and market-driven 
requirements engineering~\cite{Dahlstedt,Karlsson2007}, it promotes a 
consistent view on software engineering, and is well recognized within software 
engineering community~\cite{Sicilia1990,BudgenDavidTurnerMarkBreretonPearlKitchenham2008}.

Other frameworks, such as CMMI~\cite{CMMIProductTeam2010} and 
SPICE~\cite{Dorling1993}, are  oriented towards software process 
identification, assessment and improvement. 
However, as indicated by Giardino et al.~\cite{Giardino2014} and 
Sutton~\cite{Sutton2000}, start-up organizations are very immature in a process 
sense, thus identification of engineering processes could be difficult.

Several business practice frameworks, such as ABPMP BMP CBOK~\cite{Abpmp2009}, 
papers by Osterwalder~\cite{Osterwalder2005} and Zachman~\cite{Zachman}, aim to 
map and categorize business practices, however none of them is specific to 
start-up companies~\cite{Blank2013}. Due to youth and immaturity of start-up 
companies, traditional business processes are difficult to identify. Therefore, 
we created a simplified business practice taxonomy based on ABPMP BMP CBOK. The 
exact categories of the taxonomy are provided in the supplementary material\footnote{http://eriksklotins.lv/files/exp-reports-study-supplemental-material.pdf}.

We use the unified SWBP taxonomy to identify and categorize all relevant 
practices to explore software-intensive 
product development in start-ups.

\section{Research methodology} \label{sec_rm}

\subsection{Research questions}
Our research goal is to investigate how software engineering is practised in 
start-up companies and how software engineering contributes to other areas of a 
start-up, such as business and marketing. 
To break down the research goal further, we formulate the following research 
questions:

RQ1: What software engineering knowledge areas do software start-up 
companies consider most relevant?

We explore what knowledge areas and specific categories in software 
engineering  are of interest for start-up companies. This enables to focus 
further research on the areas that have the potential to actually support 
software start-up companies.

RQ2: How are the identified knowledge areas applied in start-ups?

We explore how the identified knowledge areas (RQ1) are implemented in 
start-up companies. This increases the understanding of how, or whether at all, 
software engineering knowledge and practices are applied.

RQ3: What are the relationships between different knowledge areas?

We explore the relationships between the applied practices to 
understand how practices and knowledge areas affect each other. The 
understanding of the relationships enables the analysis of how significant each practice is in context of other practices.

RQ4: What practices are missing to support software engineering in start-up companies?

Having identified relevant knowledge areas (RQ1), how the knowledge areas are implemented in terms of practices (RQ2), and relationships between the practices (RQ3) we identify gaps in applied practices.  We review related work to identify practices that could be useful for adaptation in start-ups.

\subsection{Data sources and collection}

We use a collection of start-up experience reports~\cite{cbinsights.com2015} as the data source. The reports represent a primary data source, i.e. we perform original analysis of data, and we perform third degree analysis, i.e. we independently analyze artifacts that are already available~\cite{Lethbridge}.

The experience reports 
describe lessons-learned from start-up companies, written by one of the 
participants after critical events occurred in the start-up, for example, a key person leaving a company, a product launch, a buyout or closure of the company. Even though the reports represent 
"multi-vocal literature", i.e. are unstructured and vary in length, focus and 
style~\cite{OgawaR.T.Malen1991,Tom2013}, they provide a rich insight on how 
start-ups perceive and approach software-intensive product engineering~\cite{Garousi2016}. The 
reports cover a much broader scope than just product engineering thus providing 
insights how product engineering is influenced by and what influence 
engineering has on other factors in start-ups.

We screened the collection to remove inaccesible or otherwise irrelevant reports, 
prior to the detailed analysis, according to the following criteria:

\begin{enumerate}
\item The report is inaccessible, for example, the link provided in the website to the actual report is broken.
\item The report clearly does not describe experience from a start-up company, for example, the report describes experiences from an established company.
\item The report clearly does not describe experiences from a software start-up, i.e. the start-up does not do software engineering in any way.
\end{enumerate}

From the initial set of 93 reports, 5 reports were removed and 88 remained for further analysis. Secondary data sources are start-up profiles~\cite{crunchbase.com2015} that record company track length, geographical location 
and members of their founding teams. Figure~\ref{rm_overview} illustrates how each data source contributed to 
answering the stated research questions.

We created a simple database tool to import, store and maintain traceability 
between different pieces of data. Prior to storing, we trimmed irrelevant 
information, i.e. ads, unrelated pictures, links, from the reports and split 
the reports into chunks by a paragraph for further analysis. A paragraph was selected as unit of analysis as it is large enough to convey a complete statement and small enough to maintain traceability. The splitting was first done with an automated tool and later manually revised. The final chunks were 1 - 16 sentences long.

\subsection{Analysis design and execution}\label{sec:design_and_execution}

\begin{figure}[ht!]
\begin{center}
\includegraphics[width=0.98\columnwidth]{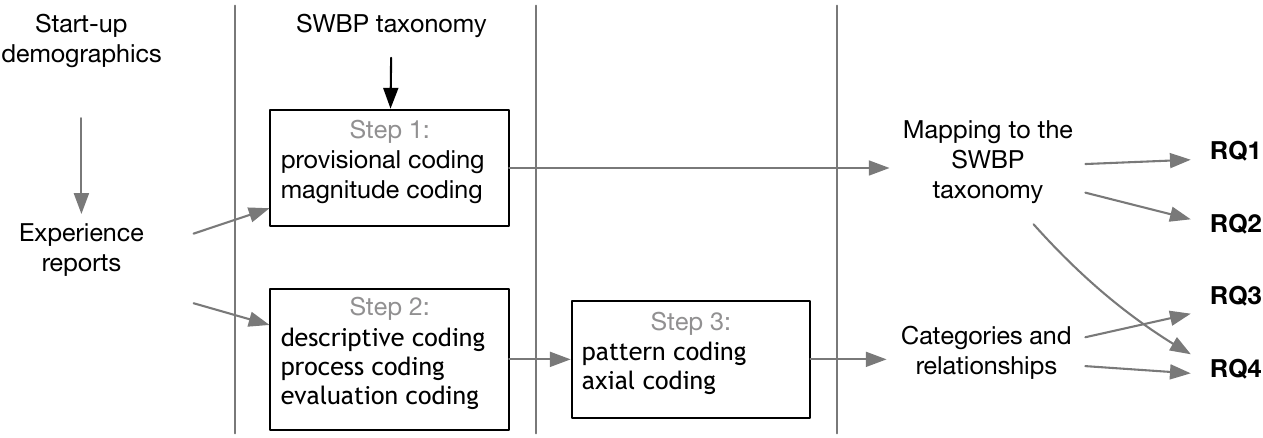}
\caption{{\label{rm_overview} Overview of the coding process and research questions%
}}
\end{center}
\end{figure}

To analyze the reports we use qualitative data analysis 
methods, that is, different types of coding and themeing of concepts~\cite{Seaman1999,Saldana2010}. Figure~\ref{rm_overview} illustrates the used data 
sources, the three analysis steps, the output and how it answers the research 
questions.

In step 1 we apply provisional coding we explore the data and identify and categorize statements 
in the reports related to software engineering or business development~\cite{Saldana2010}.
We associate statements in the reports with preformed codes from the SWBP taxonomy (see 
Section~\ref{sec_swbp}), as suggested by Seaman et al.~\cite{Seaman1999}. 

Our study is specifically angled towards software engineering practices and their 
connections to business knowledge areas, thus the SWBP taxonomy supports 
identification of relevant data to answer our research questions.
We associate statements in the reports with a knowledge area, category or a 
sub-category from the taxonomy. More explicit statements are mapped to  
sub-categories, while more general statements are associated with a knowledge 
area in general. Mapping to the lowest level enables analysis both on knowledge 
area level and a more specific analysis of what sub-categories in a particular 
knowledge area are addressed. 
To code ambiguous statements we use simultaneous coding~\cite{Saldana2010}, 
mapping the statement to multiple categories of the SWBP taxonomy. 
To further describe each identified statement we added two magnitude 
sub-codes~\cite{Saldana2010}. The first sub-code captures the impact direction of a 
described practice. We use the values ``positive", ``unknown" and ``negative" to 
capture the report's author own reflection on the impact of a practice. 
For the second sub-code we use the values ``product", ``business" and ``both" to 
capture the affected object. The scope for each code is a sentence in a report. 
Sub-codes are added on top of each code.

In step 2 we perform another pass on each report and look at the reported 
experience as a whole to 
identify key concepts, analysis 
points, leading to gains or losses in software engineering or business 
development. Looking at the reporter's opinion on what activities had 
significant impact on software engineering, we attempt to identify contextual 
factors and activities causing the high impact situation. We make use of 
\textit{descriptive} (to summarize), \textit{process} (to capture ongoing action) and \textit{evaluation}
(to assess the situation) coding~\cite{Saldana2010} jointly to capture analysis 
points in a report. Through analysis of the described situation we 
aim to differentiate between reported symptoms (e.g. running out of resources) 
and actual causes (e.g. poor resource planning due to lack of experience). Coding in this step is open, i.e. we let codes to emerge naturally without use of the taxonomy.  Examples of coding in the steps 1-2 are provided in the supplementary material
\footnote{http://eriksklotins.lv/files/exp-reports-study-supplemental-material.pdf}.

In step 3 we apply pattern and axial coding~\cite{Saldana2010} to combine 
similar analysis points and to establish relationships between emerging 
categories. As similar activities and contextual factors recur in the data, we 
group them under a candidate category. We merge, split and update the candidate 
categories during the coding process. A category gains full category status 
when its category description allows understanding of characteristics, 
conditions, consequences and interaction of the expressed concept.
To understand how different concepts influence each other we further employ 
axial coding~\cite{Corbin1990} looking for possible causes and consequences 
across all analysis points forming each category. This enables deeper 
understanding of a concept and provides multiple explanations for its emergence 
and impact.
Field memos in a form of white board drawings, notes and mind-maps were 
created to record any discoveries in the data. 
During the analysis we kept track of what specific statements from reports 
actually support the category or the relationship and continuously update a 
category description. We use this information to further develop or discard 
patterns emerging from the data.

Saldana~\cite{Saldana2010} suggests to connect categories with underlying 
concepts by applying theming portions of data. Where possible, we associate 
the categories with practices from the the SWBP taxonomy. The association 
connects the categories to state-of-the-art enabling further elaboration and 
exploration of a category.

\subsection{Answering the research questions}
We answer RQ1 (What software engineering knowledge areas do
software start-up companies consider most relevant?) by counting how frequently 
each knowledge area is discussed in the reports. Some reports repeatedly 
address the same issue resulting in multiple identical codes and impact 
sub-codes. Such repeated codes are useful for further qualitative analysis, 
however are counted only once in the quantitative analysis in order not to 
inflate the importance of a knowledge area.
By differentiating between the reported positive or negative impact on software 
engineering or business development aspects, we identify potential inadequacies 
in the application of knowledge in that particular area, specific for the 
software startup context. The analysis of this research question is presented 
in Section~\ref{results_rq1}.

We answer RQ2 (How are the identified knowledge areas applied in start-ups?) by summarizing 
reported practices relevant to each knowledge area.

We answer RQ3 (What are the relationships between different knowledge areas?) by 
developing a graph illustrating relevant knowledge areas and their relationships. By 
looking at the
number of data-points we identify the most important relationships for further 
exploration. We discuss the relationships in context of related work.

We answer RQ4 (What other practices are missing to support software engineering in 
start-up companies?) by identifying the most interconnected categories and their 
relationships in Fig.~\ref{new_overview} (RQ3). We assess the central categories in context of 
the related categories, examine if the answer to RQ2 suggest any particular practice 
pertaining the relationship, and review related work for candidate practices.

\subsection{Validity threats }

We present four categories of validity threats as proposed by Runeson et al.~\cite{Runeson2012}.

\subsubsection{Construct validity}
Construct validity is concerned with to what extent the studied operational 
measures represent 
what the researcher is attempting to investigate~\cite{Runeson2012}.

A possible threat is that we may fail to recognize relevant practices in the reports. To address this threat we use a taxonomy to support identification and categorization of statements from the reports.

We use the SWBP taxonomy as a framework to identify and categorize statements 
from the reports.

However, the SWBP taxonomy is partly based on SWEBOK and may not up to date with emerging concepts in software engineering such as 
value based software engineering~\cite{Boehm2003,Azar2007} and market-driven 
requirements engineering~\cite{Dahlstedt,Karlsson2007}. To address the threat 
that some important aspects are missed due to the lack of a comprehensive 
taxonomy we 
use two separate coding strategies, i.e. provisional coding (based on a 
taxonomy) and analysis points (independent from any taxonomy).

Due to lack of detail or terminology differences between the reports and the 
taxonomy it could be challenging to map certain statements to specific software 
engineering practices. 
We address this threat by a) applying multiple codes to 
the same statement to capture multiple interpretations and b) mapping the 
statements to different levels of the taxonomy. As a possible consequence of multiple 
codes per statement, we may incorrectly estimate the number of statements addressing a 
particular category of the taxonomy. However, multiple codes also enable identification 
and analysis of closely related and overlapping practices.

Another threat to construct validity is a possible bias stemming from the nature of 
subjective experience reports. In their essence, the reports are self-evaluations by the 
authors. They may have overlooked their own shortcomings, e.g. a lack leadership or 
technical skills, and rationalized their experience with external 
circumstances~\cite{Pronin2002}. However, due to relatively large sample and diverse 
population we likely cover different personality types, therefore minimizing 
this 
threat~\cite{Feldt2010}. An alternative solution could be to use grounded theory~\cite{Corbin1990} and study a smaller sample in more detail, thus strengthening the internal validity at the expense of generalizability.

Even though we applied multiple remedies to address the threats that stem from 
the nature of the data we have collected, we are conservative in the 
conclusions that we draw from the data and analysis.

\subsubsection{Reliability} 
This aspect is concerned with the extent to which the results and analysis are 
independent from the specific researchers~\cite{Runeson2012}.

We address reliability by ensuring transparency and traceability throughout our 
data collection and analysis, providing a detailed description of the applied research 
method. All analyzed experience reports are provided as supplemental 
material\footnote{http://eriksklotins.lv/files/exp-reports-study-supplemental-material.pdf}.

We have kept traceability information throughout our results and analysis linking specific conclusions with supporting statements in the experience reports.

\subsubsection{Internal validity}

Internal validity is threatened when a researcher is investigating one factor affecting 
another factor and there exists a third, unknown factor, that confounds the studied 
relationship without the researchers knowledge~\cite{Runeson2012}.

A possible threat here is the single researcher bias in the coding process. To address 
this threat we applied researcher triangulation. Twice in the 
coding process, at the beginning and later in the process, selected reports 
were analyzed independently by three researchers and the results discussed to 
identify and eliminate any individual biases. In addition to the triangulation, 
intermediate results were discussed among the researchers and compared to 
state-of-the-art.

\subsubsection{External validity}

External validity concerns the ability to generalize the result of research 
efforts to industrial practice and to what extent the results are of interest outside the investigated case~\cite{Runeson2012}.

A potential threat to external validity is sampling. For our study we have used convenience sampling and  
studied a relatively large number of cases from different geographical regions, market 
segments, team composition variations, product types, and consider both successful and failed cases.

The majority of start-up cases in our sample (63 of 88, 72\%) are closed companies. As elaborated in 
Section~\ref{results_overview} and Fig.~\ref{fate}, this proportion is 
representative of the whole start-up population which is nevertheless biased 
towards failed start-ups. Therefore we avoid to present any prescriptive advice 
that derives from the studied companies.

\section{Results} \label{sec_results}

\subsection{Overview of the data set} \label{results_overview}

We have analyzed experience reports of 88 start-ups. The sample of 
companies is diverse in developed products, geographical location and 
founders experience. We attempt to estimate to what extent the studied sample is representative 
to the whole population.

\begin{figure}[ht!]
\begin{center}
\includegraphics[width=0.98\columnwidth]{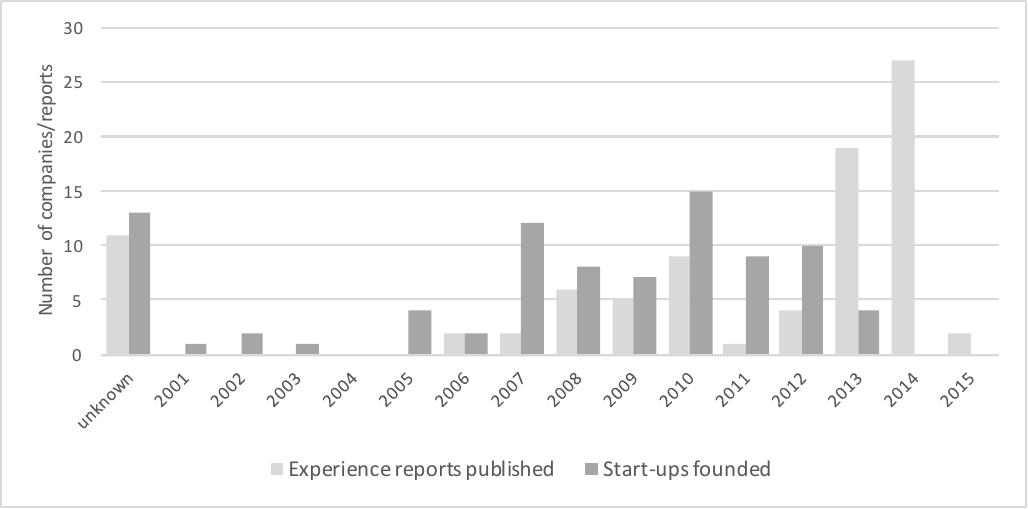}
\caption{{\label{timeline} Overview of the start-up founding and report publishing time%
}}
\end{center}
\end{figure}

Fig.~\ref{timeline} shows an overview of when companies in our sample were 
founded and when the reports were published. A total of 56 out of 88 companies 
(63\%) have provided  information on their founding and closure time. The companies were founded  between 2001 and 2013 ($Median = 2010$). For 77 companies we 
were able to identify when they have published their experience report. The 
reports were published between 2006 and 2015 ($Median = 2013$).

\begin{figure}[ht!]
\begin{center}
\includegraphics[width=0.98\columnwidth]{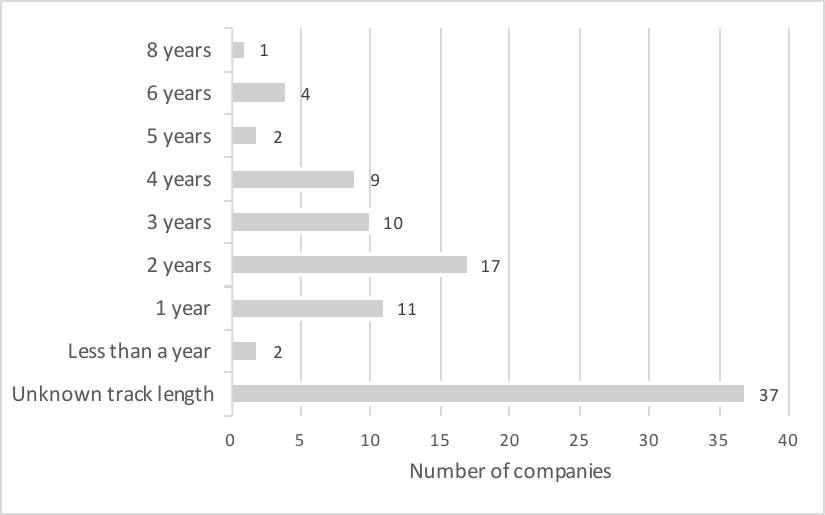}
\caption{{\label{track_length} Overview of the operational track length%
}}
\end{center}
\end{figure}

Fig.~\ref{track_length} summarizes the operational track length of the 
companies, which varies between less than a year to 8 years. 
The majority of companies for which we know the track length, 38 out of 56 
companies (68\%), have operated between one and three years.

A total of 65 out of 88 companies (73\%) have provided location information. Within this group, 48 out of 65 
companies (74\%), were located in US, 13 companies were located in Europe, and the rest were located in India, Australia and Canada.

\begin{figure}[ht!]
\begin{center}
\includegraphics[width=0.98\columnwidth]{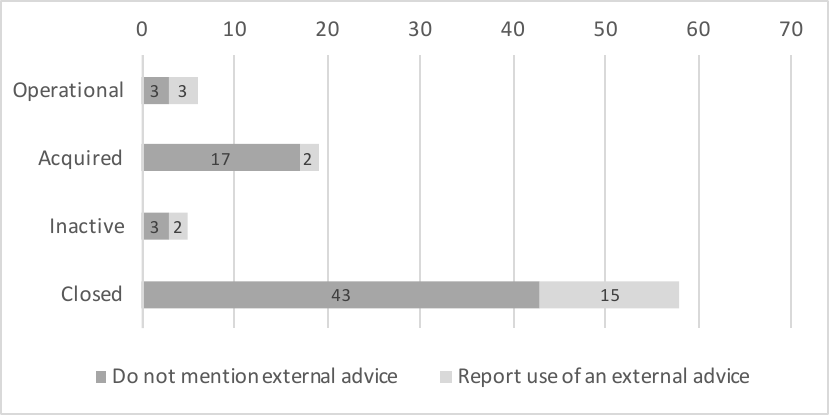}
\caption{{\label{fate} Outcomes of the companies and impact of external advice%
}}
\end{center}
\end{figure}

In Fig.~\ref{fate} we summarize status of the studied companies. In contrast to 
what is stated on the website~\cite{cbinsights.com2015}, some of the companies 
have re-emerged, continue working, or were acquired by other companies, thus 
can be considered as relative successes. Analyzing the reports and publicly 
available information, as of June 2016, we  have identified different outcomes 
of the evaluated companies. We distinguish between the
following:

\begin{itemize}
  \item Operational: The company is still in operation. This category also 
  includes companies that have re-emerged with similar products. Also, 
  companies that have pivoted, e.g. redesigned their product. Examples of such companies are: 
  \begin{itemize}
    \item GroupSpaces, available at \url{http://groupspaces.com/}
    \item Pumodo, available on iTunes as `Champion - Football Livescore, League and Cup Action'
    \item PatientCommunicator, available at \url{http://patientcommunicator.com/}
  \end{itemize}

  \item Acquired: The company, the team or its intellectual property was 
  acquired by another company. Examples of such companies are:
  \begin{itemize}
    \item  Decide, acquired by eBay in 2013
    \item  PackRat, acquired by Facebook in 2011
    \item  ReadMill, acquired by Dropbox in 2014
  \end{itemize}

  \item Inactive: The company has ceased any active sales or commercial product 
  development activities, however the product is still available to users. This 
  category also includes inactive companies that have made their products 
  available as open-source.

  \item Closed: The company has ceased any operations, the product is not 
  available to users.
\end{itemize}

In Fig.~\ref{fate} we distinguish between companies that have reported 
participation in incubator 
programs or otherwise received an external expert advice, e.g. consulted with 
investors or 
mentors. There is no clear tendency of external advice being a determinant to 
company survival, acquisition or close-down.

In the general start-up population, the failure rate is about 75\% 
\cite{StartupCompassInc.2015}. Companies that are closed down or inactive we 
consider as failed start-ups. Companies that are operational or have been 
acquired we consider successes. As Fig.~\ref{fate} shows, in our sample 6 
companies are still operational and 19 have been acquired, resulting in  a 
total of 25 (28\%) companies that can be seen as successful, while
58 (66\%) of the companies have been closed, thus can be considered as failed.
Although we do not know financial details of the operational and acquired companies, the 
percentage of failed companies in the general population (75\%), and closed and inactive 
companies in our sample (72\%) is similar.

\subsection{Knowledge Area Overview} \label{results_rq1}

As a result of provisional coding (see Step 1 in Fig.~\ref{rm_overview}), we 
identified and mapped 876 statements from the experience reports to the SWBP 
taxonomy. Saldana \cite{Saldana2010} suggests to look into how many reports mention
a particular code instead of a total count of codes in the dataset. Therefore, we 
removed identical codes per report from further frequency analysis. After this filtering our dataset contains 755 codes.

We use the number of codes to illustrate what knowledge areas and their 
subcategories are common in the reports and what rarely occurred 
\cite{Saldana2010}. There could be several explanations why a knowledge area is 
discussed in an experience 
report. One is that activities associated with a particular knowledge area were conducted 
and had some interesting effect. This explanation 
suggests that a knowledge area is relevant for start-ups either because it is useful (and 
yields positive results), or requires adaptation for use in start-ups (if the knowledge 
area was applied with good intentions but provided unanticipated results). 
Another explanation is that a company did not apply a potentially useful knowledge area, 
however reflected on their mistake in a report. This suggests that a knowledge 
area could be useful in hindsight.

\begin{figure}[ht!]
\begin{center}
\includegraphics[width=0.98\columnwidth]{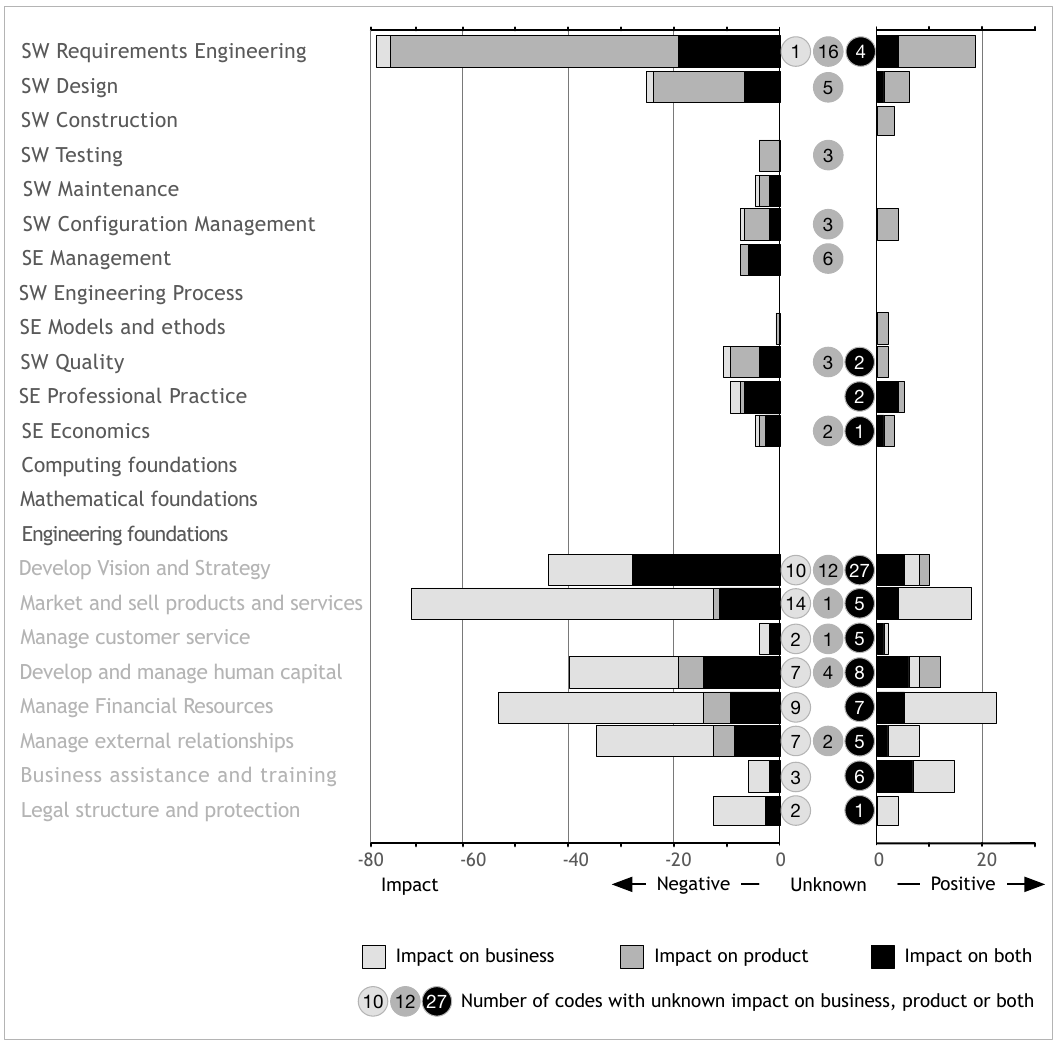}
\caption{{\label{all_ka_frequency} Overview of the number of statements associated with 
knowledge areas%
}}
\end{center}
\end{figure}

Fig.~\ref{all_ka_frequency} shows a summary of this analysis. The horizontal 
axis shows the number of reports mentioning a particular knowledge area; 
direction (positive or negative) indicates impact. Differently shaded bars show 
whether the impact is discussed as having impact on business development, 
software engineering or both. The total length of a bar indicates the total 
number of codes referring to particular knowledge area. Note, that a knowledge area could be
discussed from various aspects in a single report, thus resulting in multiple different codes per 
same knowledge area.
The number of 
statements discussing a particular knowledge area but not specifying any impact are shown 
in circles between the bars. In total, 209 (28\%) statements address software engineering 
aspects, and 296 (39\%) address business aspects of start-ups.

As Fig.~\ref{all_ka_frequency} shows, software requirements 
engineering, software design and professional practice are the top 
three most discussed knowledge areas in relation to inadequacies in product engineering.
Software engineering process, computing foundations, mathematical foundations 
and the engineering foundations knowledge areas are not 
discussed at all in the reports.

\begin{figure}[ht!]
\begin{center}
\includegraphics[width=0.98\columnwidth]{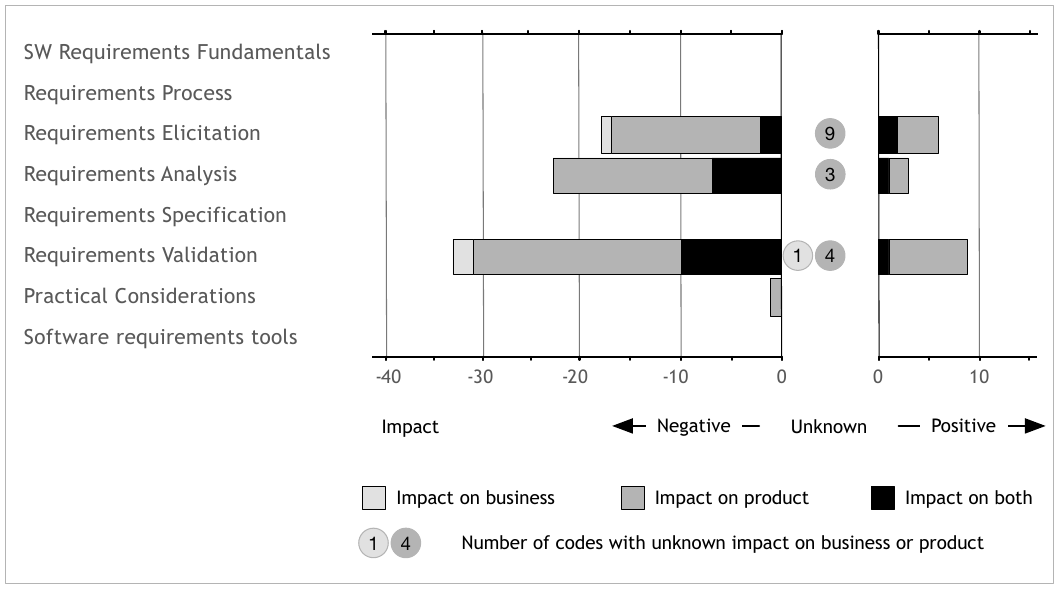}
\caption{{\label{rm_frequency} Breakdown of the software requirements engineering knowledge area%
}}
\end{center}
\end{figure}

\begin{figure}[ht!]
\begin{center}
\includegraphics[width=0.98\columnwidth]{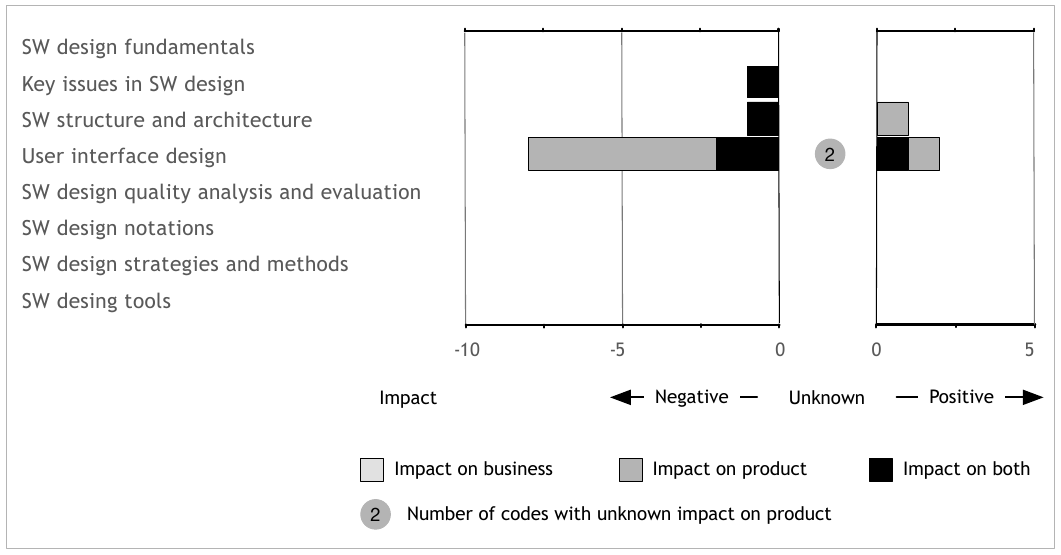}
\caption{{\label{sd_frequency} Breakdown of the software design knowledge area%
}}
\end{center}
\end{figure}

\begin{figure}[ht!]
\begin{center}
\includegraphics[width=0.98\columnwidth]{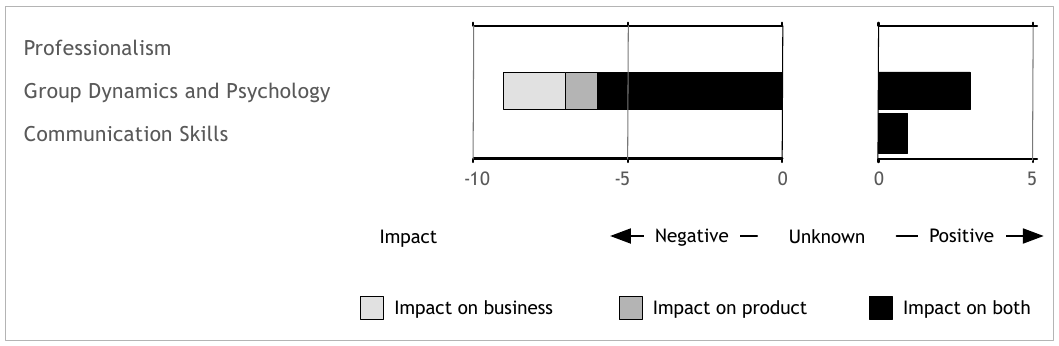}
\caption{{\label{pp_frequency} Breakdown of the software professional practice knowledge area%
}}
\end{center}
\end{figure}

Figures~\ref{rm_frequency}-\ref{pp_frequency} illustrate what specific 
subcategories of requirements engineering, software design and engineering 
professional practice are discussed in the reports. From the requirements 
engineering knowledge area, requirements validation, analysis and elicitation 
are the most discussed sub-categories. From the software design knowledge area, 
the most discussed sub-category is user interface design. From the software 
professional practice the most discussed sub-category is group dynamics and 
psychology.

\section{Analysis and discussion} \label{sec_analysis}

In this section we synthesize our analysis into the software engineering knowledge areas 
reported by startups to answer the remaining three research questions. First, we report how a knowledge area is 
applied, what specific practices are mentioned and what specific challenges are 
discussed in the reports in relation to the knowledge area (RQ2). Second, we 
explore what relationships between knowledge areas are reported to understand 
how software engineering knowledge areas influence each other (RQ3). Third, 
we look into related work from similar engineering contexts, compare challenges 
between start-up and other engineering contexts, and identify potentially 
useful practices to solve engineering challenges in start-ups (RQ4).

\begin{figure}[ht!]
\begin{center}
\includegraphics[width=0.98\columnwidth]{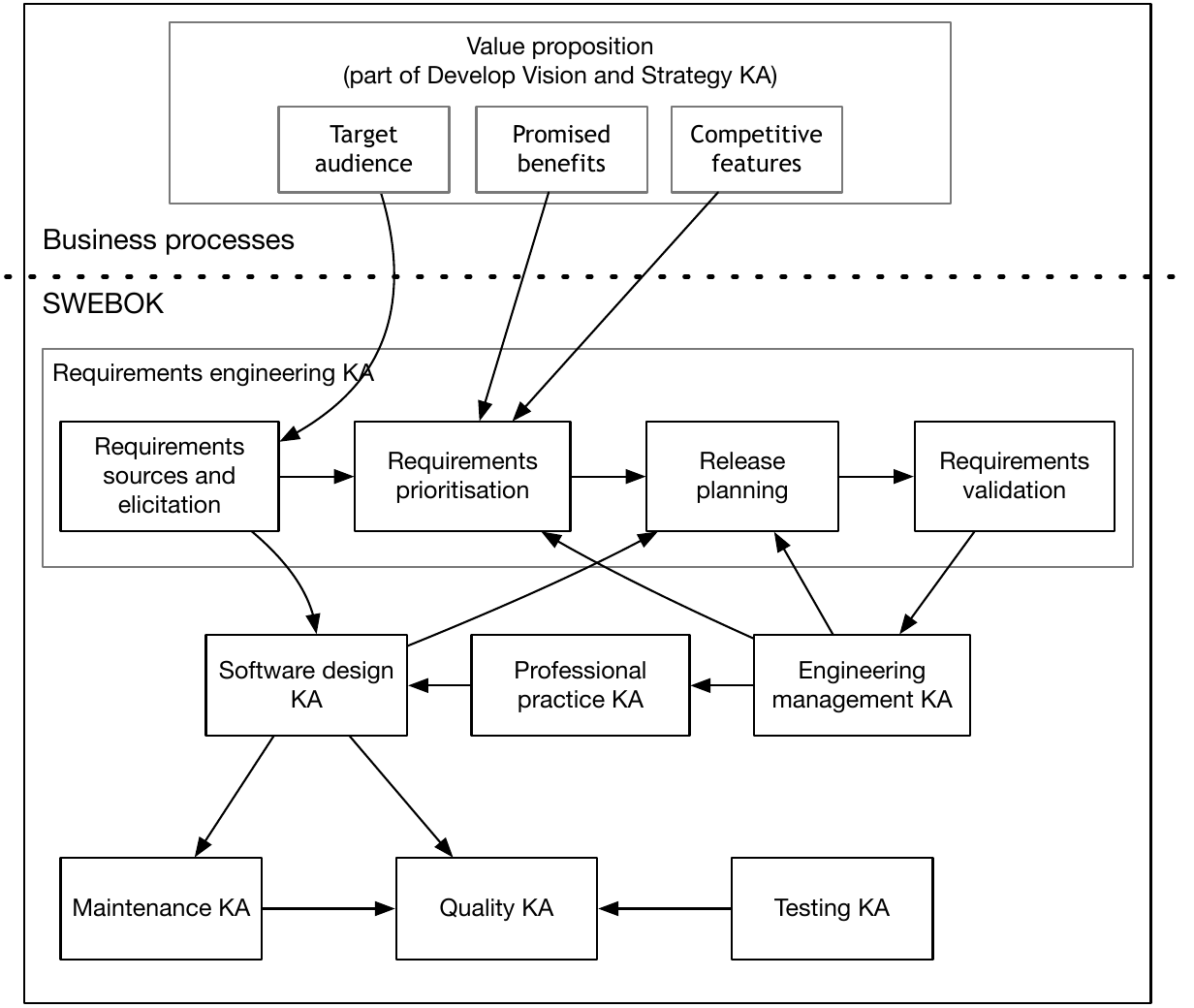}
\caption{{\label{new_overview} Software engineering categories and their relationships%
}}
\end{center}
\end{figure}

Fig.~\ref{new_overview} shows key knowledge areas and their relationships of 
software-intensive product engineering in start-up companies (identified by 
applying pattern and axial coding as explained in 
Section~\ref{sec:design_and_execution}). In Fig.~\ref{new_overview}, boxes 
represent knowledge areas, arrows denote a relationship between knowledge 
areas. A relationship indicates that a parent category provides input, i.e. 
information, to a child category. 
In further subsections we discuss the knowledge areas and their relationships in detail.

To support traceability between our analysis and data in the reports, we use 
references to the original data. The references are represented by identifiers 
in curly braces after a statement, formatted in the following way: C\textless 
\{\textit{company\#}\textgreater-\textless\textit{chunk\#}\textgreater\}. The 
identifiers refer to the supplementary material available 
online\footnote{http://eriksklotins.lv/files/exp-reports-study-supplemental-material.pdf}.

\subsection{Develop vision and strategy knowledge area}

Our analysis shows the process of identifying a product value proposition as a bridge between marketing and engineering aspects of a product. Value proposition provides an essential input for starting  software requirements 
engineering activities. The value proposition is a structured description of a 
product idea. It outlines what is the target audience for the product is, what 
benefits the product aims to deliver, and what the competitive features of the 
product are~\cite{Carlson2006}.

The reports discuss identification of the value proposition as an iterative process where the initial formulation is brainstormed, and then improved by means of market research, customer interviews, prototype demonstrations and similar activities 
\{C2-50,
C11-10,
C24-3,
C29-63,
C37-30,
C48-5,
C50-9,
C52-26,
C63-45,
C64-33,
C67-45\}. 
The value proposition bridges the gap between market research (with a goal to 
explore market potential of the product) and requirements engineering (with a 
goal to identify a feasible solution)~\cite{Hague2004,Dahlstedt}.

As shown in Fig.~\ref{new_overview} a structured formulation of the product idea helps to identify specific goal 
level requirements which then are broken down into more specific functional and 
quality requirements by requirements engineering activities. Target audiences 
help to identify stakeholders for requirements elicitation activities. Inadequacies in the value proposition may hinder 
requirements engineering activities. For example, an unclear overall product 
goal makes it difficult to specify criteria for requirements prioritization, 
release scoping and for identifying stakeholders.

\subsection{Requirements engineering knowledge area} \label{sec:summary_RE}
Software requirements engineering is a set of activities to capture the needs 
and constraints placed on a software product, and to identify a feasible 
solution that contribute to solving a real-life problem. Therefore,  
requirements engineering can take both problem and solution oriented 
view~\cite{Society}.

As shown in Fig.~\ref{all_ka_frequency}, requirements engineering is the most discussed software engineering knowledge 
area in the reports. Further analysis of statements from the reports, 
illustrated in Fig.~\ref{new_overview}, suggests that requirements engineering 
is the central software engineering activity in start-ups.

Start-up companies operate in a market-driven environment, thus initial 
requirements are invented by a start-up team~\cite{Dahlstedt,Ambler2002}. In similar contexts outside start-ups, requirements are validated by internal feasibility reviews, interviews, 
surveys, crowd-funding success and other techniques that are applicable in the
pre-development stage~\cite{Fabijan2012,Ambler2002}. Requirements negotiation 
takes place to prioritize what features to implement next~\cite{Tingling2007}.

The experience reports suggest that start-ups use a similar approach to requirements engineering. Software is developed in short iterations aimed to implement and validate a slice of requirements. Results from the validation are used as input for subsequent iterations. As Company \#1 
reflects on quickly building a prototype, testing it and only then undertaking more extensive mobile application development: 

\myquote{``We had a mobile website prototype in front of users within a week and 
iterated based on that before building out the native [mobile application] version.''}

Requirements engineering drives the software development process by helping to acquire domain knowledge, explore problem domain, and to identify potential solutions~\cite{Hofmann2001,Society}. As 
put by Company \#66: \myquote{``One of the key lessons I learned is that great startups 
have a blindingly obvious, ideally really large and painful problem that the company is 
trying to solve. Solving this problem should drive almost every decision in the 
startup.''} Exploring the problem domain and user needs is one of the key practices in 
early stage start-ups \cite{Crowne2002,NeilC.Churchill1983}. Our findings are consistent with Hofmann et al.~\cite{Hofmann2001} who argue 
that inadequacies in requirements engineering are the single largest cause of software 
project failure.

In the following subsections we discuss sub-categories of the requirements engineering 
knowledge area.

\subsubsection{Requirements sources and elicitation}

This category represents practices to collect requirements and to identify
sources from where engineers can collect 
requirements~\cite{Society,Dahlstedt}.

The reports suggest that start-ups operate in a market-driven context and that 
the initial requirements are derived from the product value proposition. 
Interviews, surveys, observations and demonstration of prototypes are reported 
as methods to adjust goals, discovering new requirements, and to validate existing 
requirements \{C1-45,
C14-20,
C29-2,
C48-5,
C50-9,
C59-24,
C75-4,
C79-102,
C86-10\}

The data from the experience reports suggests also that start-up teams use 
local businesses, people from their social network and even their teams as 
requirements sources. Similar products, industry standards and regulations, and 
partnership agreements are discussed as important requirements sources in 
addition to customer feedback. Examination of similar products is reported as 
useful to identify base functionality of a product and to spot opportunities 
for innovation \{C06-12,  C1-45,
C14-8,
C14-20,
C33-73,
C39-10,
C50-8,
C61-43,
C69-6,
C75-4,
C79-102,
C86-10\}.

The reports suggest that the utilization of customer feedback depends on 
access to requirements sources and interviewer's skill to discover actual customer 
needs. The access could be limited by, for example, physical distance and 
inadequacies in identifying potential customers. Mistaking curious people for 
potential customers can lead to false requirements hindering the product's 
market potential. Some companies report testing customer interest by asking for 
an upfront payment 
\{C14-34, C21-7, C02-25\}. As Company \#14 states:

\myquote{ ``I think we did not 
understand that the real purpose of selling was validation (or invalidation) 
and had the 'always be closing' mindset at a too early stage of the company. 
Later, I have been joking that during the validation process, if customers 
don't buy, you should open a champagne bottle and celebrate that you found one 
way that didn't work and are now a lot closer to success.''}

The start-up companies reflect on the importance of early customer feedback and the 
dangers of not using customer input in the requirements engineering process. Even 
though gathering of customer input is discussed as difficult due to a physical 
distance and vague understanding of the target market, customer input is reported 
as an essential part of requirements engineering. Companies that have neglected 
early customer feedback report poor product reception in the market and wasted 
resources on developing unwanted features, often leading to the company's 
collapse \{C03-10, 
C22-4, 
C35-23, 
C50-8, 
C52-20, 
C34-2, 
C59-24, 
C75-4, 
C75-17, 
C75-19, 
C76-11, 
C78-15, 
C78-19, 
C86-10, 
C88-4\}.

A commonly reported difficulty is to create an engaged community of early 
customers of the product. This community facilitates requirements elicitation, 
validation and other activities were direct customer feedback is essential. The 
reports suggest that initially a person may show genuine interest in the 
product, however, if the product does not 
solve an actual problem for the customer, the interest fades away quickly
\{C06-6, 
C35-20, 
C52-22, 
C59-7, 
C65-16, 
C67-38, 
C69-6, 
C82-16, 
C83-6\}.

The reports suggest that misuse of customer feedback stems from difficulties to 
identify and access requirements sources, i.e. customers, and poor elicitation 
methods, for example asking the wrong questions \{C4-22,
C4-72,
C59-7,
C59-24,
C61-43,
C79-102,
C83-6\}. As Company \#2 reflects: \myquote{``People 
compliment you on the idea because they believe it will be so useful for people 
other than themselves. i.e., they get into advisor mode.''}

\paragraph{Discussion:}
As shown in Fig.~\ref{new_overview}, using value proposition to identify concrete requirements sources and software requirements is one of the first steps in product engineering activities. Inadequacies in value proposition and requirements engineering activities could hinder any further engineering activities. Unclear quality and functional 
requirements lead to over or under-engineering of the product.

Identification and access to useful requirements sources is essential for 
requirements elicitation \cite{Mitroff1983}. In a market-driven context, a 
company must solve the practical problem on how to select a manageable number 
of users, e.g. early customers, to perform requirements elicitation 
activities.

Pacheco et al. \cite{Pacheco2012} suggest to classify all likely users
and to study all of the user classes to identify their role in the 
product. Pruitt et al. \cite{Pruitt2003} suggest that the use of superficial 
characters representing users of the product, i.e. 'personas', helps to 
identify different user groups and to facilitate discussion around the 
requirements. The personas 
could be created with help of a small group of customers or domain experts and further detailed with interviews, 
surveys and ethnographies to create more detailed descriptions of the 
users and their needs 
\cite{Miller2006,Pruitt2003}. This lightweight practice could be useful for 
start-ups when actual customers are not readily available.

Fabijan et al. \cite{Fabijan2012} suggest
different customer feedback collection techniques useful at different 
development stages. Since access to actual users for face-to-face 
interviews is usually limited, start-ups could use indirect requirement sources 
such as listing the product idea on a crowd-funding website 
\cite{Fabijan2012,Pruitt2003}, validating the product idea and discover new 
requirements with less effort.

Due to practical restrictions, only a limited number of potential users can 
be involved in elicitation and any requirements are generalized over 
a larger population. However, such approach poses risks of biases, such as sampling (e.g. consulting only expert users as requirements sources), and data collection method (e.g. utilizing only quantitative surveys). 
Wilson~\cite{wilson2006triangulation} argues that triangulation and use of multiple methods, measures and approaches must be explicitly interweaves in requirements elicitation process. He argues, that the best results can be achieved by mixing qualitative and quantitative methods, and using multiple complimentary data sources.


Karlsson et al. \cite{Karlsson2007} report that technology focused companies often neglect user feedback in favor of inventing requirements internally. This is partly due to difficulties obtaining feedback on a new product that is unknown for a market, and partly due to focus on technology rather than actual customer needs \cite{Karlsson2007}. 
The reports suggest that start-ups often use interviews to elicit requirements 
from users, however users are not always able to articulate their needs. Davis 
et al. \cite{Davis2006a} identify four typical situations in requirements 
elicitation and argue that each requires specific elicitation techniques. For 
example, if a user and the analyst share knowledge about a specific 
requirement, simple questioning to verify the requirement could be sufficient. 
However, if a requirement is unknown to both sides then mutual exploration of 
the problem and requirements discovery are a more suitable approach to 
elicitation~\cite{Davis2006}. This resonates very well with findings by 
Kujala \cite{Kujala2008} arguing that it is beneficial to empower and involve a group of key customers in daily development activities.

\subsubsection{Requirements Prioritisation}
Requirements prioritisation is a requirements analysis activity to categorize requirements by how essential they are for meeting overall goals of the product. The requirement priorities need to be balanced against resources, time and other constraints~\cite{Society}.

Requirements prioritisation is discussed most commonly in relation to 
identifying features for the smallest viable feature set, i.e. a minimum viable 
product (MVP) \cite{Junk2000}. The MVP is reported as useful to showcase the 
main advantages of the product to users and to spot inadequacies in product 
features or design early \{C75-17, C14-11\}.

The reports indicate that customers, own ideas, competitors and similar sources 
provide a constant flow of ideas for new features and improvements. However, 
due to resource limitations, only a few can be implemented. Start-ups report on 
selecting features that deliver the most value to their customers. However, 
this process is reported as difficult without mentioning any specific practices
\{12-110,
14-27,
15-25,
33-6,
43-6,
48-9,
50-9,
52-6\}.

Requirements prioritization is reported as challenging, specifically the 
selection of prioritization criteria. To maintain a product focus and to stay 
within resource, time and quality constraints, the company must prioritize what 
features are the most relevant to deliver a promised value proposition 
\{C50-11, C50-11, C57-12, C71-24\}.

Some companies reflect that their challenges with requirements prioritization 
originate from a vague value proposition, i.e. unclear product goals and 
benefits. The reports suggest that consequences of poor requirements 
prioritization 
are over-scoped product releases and wasted resources on implementing unwanted 
features \{C76-5, 50-18, 69-14, 79-48\}.

\paragraph{Discussion:}

As illustrated in Fig.~\ref{new_overview}, requirements prioritization goals 
are defined by the product value proposition.

Quantifying value is a complex task and often involves making a compromise between interests of different stakeholders. When maximizing value is used as a prioritization goal, different perspectives of value need to be considered. Khurum et al.~\cite{Khurum2012} propose a breakdown of software value aspects therefore enabling discussion about different perspectives on value.

Lethola et al. \cite{Lehtola2005} identifies a need for alignment between 
business and engineering activities in a market-driven setting. The authors 
discuss use of roadmapping as a technique to align product and market 
perspectives. A road-map helps to connect immediate engineering goals with 
higher level objectives and to facilitate the discussion between different 
stakeholder perspectives, i.e. customers, business and engineering.

\subsubsection{Release Planning}

Release planning is closely related to requirements prioritisation and concerns 
the identification of sets of requirements that can be delivered to customers 
and provide competitiveness in the market~\cite{Carlshamre2000}. In a 
market-driven setting there is a constant pressure to deliver features 
faster~\cite{Giardino2014}. However practical challenges, such as requirements 
interdependencies, need to be resolved.

When planning product releases, start-ups follow two general approaches: 
frequently releasing small increments and delivery of a fully-fledged product. 
The frequent delivery approach starts by creating a very simple functionality, 
even mock-ups, and continues until the product matures. A continuous delivery 
process allows to conduct continuous requirement validation and to immediately 
adjust the product direction \{C33-33, 
C48-4, 
C53-46, 
C54-9, 
C54-11, 
C87-60, 
C46-13\}. Fully-fledged releases take more time to build, thus continuous validation of the product direction is challenging. 
Moreover, as validation takes place after the release, substantial effort is 
put on risk to be wasted \{C14-11, 
C14-42, 
C52-20, 
C78-19, 
C78-20, 
C87-60, 
C46-13\}. 
Attempts to launch a fully-fledged version are most commonly discussed in 
relation to neglect of user input and focus on technology rather than an actual 
customer need \{C14-11, 
C14-42, 
C52-20, 
C78-19, 
C78-20, 
C87-60, 
C46-13\}.

Due to market pressure or internal uncertainty of what customers expect from the 
product, companies desire to satisfy customers with a more complete and 
polished 
product. However, implementation of more features or higher quality requires 
more resources and postpones the opportunity to demonstrate the product to 
users, thus hindering requirements validation activities. Companies that have 
leveraged on early user feedback and have launched a less complete product,  
report fewer difficulties in marketing the product \{C35-23, 
C50-11, 
C52-20, 
C57-12, 
C71-24, 
C75-17, 
C75-18, 
C76-5, 
C78-19, 
C82-12, 
C86-11\}.

We found that companies often
overscope their releases aiming to deliver a more ``ground-breaking'' product 
in hopes for more positive user feedback \{C14-11,
C52-20,
C46-13,
C78-19,
C87-60\}. As Company 
\#59 states: \myquote{``We should have concentrated on the core idea and 
launched a Minimum Viable Product (MVP) to test the concept, as we initially 
had planned even though we never had heard of the concept of an MVP. We kept 
building more features, since we always felt that 'the service needs X because 
Flickr has it too' or 'he/she said he needs that feature'.''} 
Overscoping could be a consequence of poor requirements prioritization.

\paragraph{Discussion:}
As illustrated in Fig.~\ref{new_overview}, release planning is closely related to requirements prioritisation and 
requirements validation. Prioritization provides 
means for identifying requirements to be included in a product release. 
Requirements in the release are demonstrated to customers and, thereafter, 
validated by customer feedback.

Bjarnason et al. \cite{Bjarnason2010} recognize that scoping of product 
releases is challenging. They report that an unclear vision of overall goals, 
constant inflow of requirements, and miscommunication are some of the reasons 
for over-scoping the releases \cite{Bjarnason2010}. As shown by the experience 
reports and supported by Bjarnason et al. \cite{Bjarnason2010}, consequences of 
over-scoped releases are unmet customer expectations, wasted effort and delays.

Dahlsted et al. \cite{Dahlstedt} and Alves et al. \cite{Alves2006} report that 
in market-driven requirements engineering most requirements validation takes 
place after the product is released to the users. Therefore, frequent releases 
enables early identification of potential flaws in the value proposition or the requirements.

Incremental delivery of the product and frequent adjustment of plans are 
described as key practices of Scrum \cite{Rising2000}. Rising et al. 
\cite{Rising2000} reports that organizing development in sprints and 
prioritizing features for upcoming release helps to deal with uncertainty and 
changing requirements. Moreover, the Scrum method implies that after each 
iteration an assessment of progress, user feedback and re-prioritization of 
tasks takes place. Such rigorous approach to development and planning helps 
to break down the product to manageable chunks and progress is made even if 
requirements change. Predictable timing and scope of product releases 
encourages users to adopt the product \cite{Rising2000}.

\subsubsection{Requirements Validation}

Requirements validation covers practices to ensure that engineers have understood the requirements and the proposed solution actually solves the original problem~\cite{Society}.

The reports suggest that start-up companies aim to focus their activities around continuous requirements 
validation. The most commonly discussed technique is to implement requirements 
in a early version of a product, i.e. a prototype, demonstrate it to the users 
and to collect feedback, commonly called a feedback loop \{C01-47, 
C14-8, 
C29-3, 
C35-20, 
C52-22, 
C54-9, 
C55-16, 
C61-7, 
C71-23, 
C75-18, 
C86-14, 
C87-56, 
C34-4, 
C46-13\}.

User feedback is used both to validate the requirements and to identify new 
user requirements for the product. In addition, interviews with users are 
reported as useful to review and discuss the requirements before prototyping 
\{C02-11, 
C06-6, 
C14-8, 
C29-33, 
C63-13, 
C34-2\}.

The companies report on using various metrics to gather quantitative data how customers use the product. The collected metrics are used to validate 
requirements and to steer further product development \{C1-66,
C48-5,
C50-18,
C53-72,
C57-12,
C64-17,
C75-10\}.

However, many companies 
have failed to establish the feedback loop either due to the lack of an 
internal engineering process to manage 
user feedback or the difficult access to users \{C49-21,
C50-18,
C34-2,
C86-10\}. As Company \#14, building a software tool for ordering photo prints on-line, reflects: \myquote{``Iterations took longer than 
planned for us, because small print labs were often quite busy and did not have 
time to immediately have a look at the new version and give feedback. [..] When 
they finally had time to try out the new version, if they felt that it still 
needed improvement or they came up with a new feature that would be needed, the 
launch was likely to be postponed by at least a month.''}

\paragraph{Discussion:}
As shown in Fig.~\ref{new_overview}, requirements validation in start-ups is closely related to release planning and provides an input to planning activities. Release planning determines what features are released and, therefore, undergo validation. Outcomes from the validation are used to adjust further product direction.

The most recurring issues in requirements 
validation are the lack of a structured process to utilize user feedback and 
the inability to select relevant metrics. However, the experience reports offer 
little details on specific practices addressing these issues. Our findings are 
consistent with Olsson et al. \cite{Olsson2015} who identify similar issues in 
established companies developing software-intensive products.

Hanssen et al. \cite{Hanssen2006} report on involving expert users in a 
deploy-test-evaluate loop. The expert users are central in testing and 
evaluating each release. However, the authors also emphasize the required 
overhead to maintain the user-developer relationship, to keep the users engaged 
and to make strategic decisions on the product direction.

Further research is required to understand how to identify users to be involved 
in development process and to what extent methods by Hanssen et al. 
\cite{Hanssen2006} could be applied in start-ups.

\subsection{Software design knowledge area} \label{rq2_sw_design_ka}

Software design is a set of activities and a result of defining software architecture, components, interfaces and
other characteristics of the system, supporting its construction. Software 
design can take place before the construction process as in plan-driven 
contexts, or interweave with the construction process as in an agile 
setting~\cite{Society,Yang2016}. 

As shown in Fig.~\ref{sd_frequency}, most (22 out of 34) statements associated with the Software Design knowledge 
area lack details for mapping to subcategories. The remainder of the statements 
specifically discuss the User Interface Design subcategory.

The reports offer very little information on the actual construction of the 
product, coding or integration of components. Instead, the reports discuss 
design decisions behind selecting one or another construction technology, 
components or design goals.

Statements from the reports suggest that start-ups aim to release their products or services fast, thus spending little time on upfront software design. Start-ups opt for incremental designs and faster product releases \{47, 3101, 3956\}. Scalability and flexibility of the product are identified as primary goals of software design \{264, 1245, 1836, 1922, 2982, 3249, 
3956, 1305, 733,1166\}.

Start-ups 
report on attempts to leverage on cutting-edge technologies with the aim to gain a
competitive advantage such as faster 
time-to-market or additional features. However, new technologies are often 
reported as immature causing product quality issues. As Company \#3 states 
\myquote{
``Sure, it was seven years ago, pre-iPhone and pre-Android, so it was ahead of its time, we had to use Adobe Flash on a browser which sucked in so many ways I can't even start to explain how bad it was. Technology would be so much better and more important all mobile today.''}

Selection of technology also concerns third-party solutions that can be 
integrated and configured to constitute the product. Third-party components are 
used as a method to deliver functionality with little development effort. 
Leveraging on existing functionality of third-party components is reported as a 
key practice in software design. Some companies that have not leveraged on 
third-party components admit lack of skill and experience in software design \{C04-23, C69-12\}.

Several reports mention good product user experience as an important quality and their efforts to improve it. However, no specific practices regarding user experience engineering are mentioned \{6-11, 14-15, 35-17, 66-35, 63-54\}. User interface design is recognized as having an impact on customer behavior 
and attitude towards the product. As Company \#66 reflects on user interface and user 
experience design: \myquote{``The team never properly sat down and brainstormed the UX. 
Quick decisions were made to get the MVP out the door and these had serious impacts on 
how the product was received by customers.''} Constantinides et 
al.~\cite{Constantinides2010} and  May et al.~\cite{May2012} also recognize the 
importance 
of user interface and its impact on product adaptation.

Other goals of user 
interface design are the development of a product's visual appeal, to establish 
a brand identity, to gain attention from media \{C02-87, 
C33-36, 
C67-13\}, search 
engine optimization \{C17-16\}, and promoting viral effects in social networks 
\{C66-35\}. 
An iterative approach of frequently updating the user interface 
and measuring changes in the user behavior is reported as a viable practice to build user interface of a product
\{C01-47, C02-80, c12-126, C14-33, C33-33, C52-19, C82-13, C46-13, C54-29, C63-06, C02-81\}.

The reports suggest that start-up companies use brainstorming \{C66-42\}, mock-ups and wire-frames \{C14-15\} to design user interfaces of a product. Frequent iterations \{C33-36\}, experiments \{C66-35\} and usability tests \{C14-15\} are applied to continuously improve the user interface \{C02-87, 
C35-36, C61-17, C66-35\}.

However, under a tight schedule, the process could be abandoned and user interface designs are done in a hurry with little consideration \{C14-15\} causing quality issues later. Attempts to salvage a product that is unsuccessful for other reasons by tweaking the user interface leads to wasted resources with little or no gain \{C02-87, 
C14-15, 
C63-54\}.

\paragraph{Discussion:}
The reports suggest that when the understanding of requirements is vague, it is useful to put together 
a quick prototype demonstrating a feature in question. The prototype is used to 
gather user feedback before any extensive development takes place \{C33-14, 
C57-26, 
C71-18, 
C78-25, 
C79-50, 
C79-115\}. However, sticking more features into a makeshift product degrades 
the architecture and technical debt accumulates over time. Maintaining a 
manageable level of technical debt and creating an architecture supporting 
changing requirements and enabling quick prototyping is a major challenge 
\{C04-58, C14-12, C22-04, C33-42, C50-11, C58-15, C64-31\}.

Software design activities in start-ups are closely related to requirements 
engineering.  Non-functional requirements determine the required level of quality as an input 
for software design activities. However, vague or invented non-functional 
requirements could lead to under or over-engineering of a product 
\{C04-22,C04-71, C23-3, C75-23, C55-14\}.

Quality requirements constraining internal aspects of the product, such as 
time-to-market or maintenance costs, are repoted as often overlooked. Poor or 
neglected quality requirements may create pitfalls in the long run:  
inadequately high maintenance costs when product is launched or overly long 
release cycles \{C04-22,C04-71, C23-3, C75-23, C55-14, C32-3, C78-25, C35-22, 
C79-115\}.

Creating software design that requires minimal lead time and can accommodate 
changing requirements while maintaining high product quality is a challenge. 
The reports suggest that it takes skilled engineers to build such designs and 
reflections on how inadequate engineering skills had contributed to poor design 
leading to poor product quality \{C79-50, C75-23, C53-68, C72-18, C73-40, 
C01-02, C02-80, C35-52, C43-17, C49-26, C52-19, C67-33\}.

A study by Woods~\cite{Woods2015} suggests that goals of software architects often clash with goals of agile teams, however there are simple principles that allow both to benefit from each other. For example, breaking the software into smaller components and delivering incrementally helps to avoid large upfront designs. Communicating architecture principles to the developers help the team members to understand why architectural structures exist and what are most important characteristics of the architecture. Yang et al.~\cite{Yang2016} identifies forty three software architecture approaches that can be used in an agile context. The identified approaches range from naive (considering architecture only for current iteration) to use or architectural design patterns and cost-benefit analysis.

State-of-the-art in agile software architecture offers a variety of practices and guidelines that could be relevant for start-ups. However, it needs to be explored which exact practices are most efficient to address start-up specific challenges.

\subsection{Software engineering professional practice knowledge area}\label{rq2_pp}

The software engineering professional practice knowledge area comprises skills, 
knowledge and attitudes that an engineer must posses to practice software 
engineering~\cite{Society}. The reports discuss various aspects of professional 
practice, such as decision making, motivation, trust, importance of good 
software engineering skills, and ability to learn.

Difficulties in communication are discussed as having a significant impact on 
decision making, motivation, trust and general climate in the team. As Company 
\#79 describes the communication between co-founders: \myquote {``Overall, the 
most important [challenge] is that Nathan and I had difficulty communicating in 
a way which would allow us save the company, and that this really drained out 
motivation.''}

The reports contain descriptions of team structures ranging from hierarchical 
to flat. In a start-up team the highest authority are founders. However, some 
founders empower and involve other team members in making important decisions 
while others exercise autocracy in all aspects of their companies \{C01-46, 
C12-54, C62-17, 
C74-40, 
C78-17\}.

Autocracy is discussed in the reports as a cause and consequence of lack of 
trust between team members, miscommunication of company goals and lack of 
transparency in decision making. 
However, involving the whole team in every decision hinders performance and 
team 
motivation. Separating areas of responsibility and empowering team members to 
make decisions are discussed in the reports as viable practices for decision making \{C05-06, C12-83, C16-08, C17-19, C46-05, C66-42, C69-04, C71-13, C74-40, C77-22, C78-17, C84-12\}. Company \#17 points out that in a 
dynamic start-up environment it is difficult to make decisions based on previous experience. Instead decisions should be based on data and experimentation:  
\myquote{``No one has any idea what is going to work and what's not. Don't listen to the people who think they know. Sure, this one didn't pan out, but each failure helps us navigate the thousands of decisions we will need to make for the next one. That knowledge helps us build better things that will last longer.''}

Mutual trust between team members is reported as an important factor for good 
teamwork and decision making. Founding teams with joint previous experience 
reflect on team issues more positively and reflect on mutual trust as a 
contributing factor to good teamwork. The inability to communicate mutual 
expectations, intentions and motivate own decisions hinders trust 
\{C49-26, 
C62-19, 
C72-18, 
C73-32, 
C73-34\}.

Capabilities to learn new emerging practices, adapt to an uncertain environment 
and collaborate are reported as essential in a start-up environment. Some 
reports discuss how unanticipated personality traits have contributed to 
team break-up and company collapse, suggesting that good team 
composition is essential in start-ups  \{C02-47, 
C14-2, 
C21-17, 
C29-4, 
C73-32, 
C46-5\}.

A working environment encouraging communication and collaboration, such as a dedicated office space and joint activities, boosts performance and increases motivation \{C56-16, 
C61-65, 
C71-14, 
C73-9, 
C73-34, 
C77-22\}. Working remotely is reported to have negative effects in the long term \{C21-14, 
C75-15, 
C77-3\}, however, when done with consideration, working remotely can have positive effects, i.e. to avoid disturbances in the office \{C75-12\} or being closer to the target market 
\{C01-45, C21-14\}.

The reports discuss 
how initial optimism for fast success vanishes and development tasks shift from 
inventing the product to less exciting activities such as handling customer 
service \{C02-50, 
C02-74, 
C04-20, 
C04-22, 
C77-3\}. 
When a critical motivator is not present anymore, a team member may leave the company or start following his own agenda \{C01-28, 
C12-142, 
C21-16\}.
Motivation to work in a start-up is often discussed in the reports in relation 
to shared goals and vision. A lack of shared understanding about changing goals 
is reported 
as a consequence of poor value propositions \{C01-58, C12-142, C22-07, C53-86, 
C71-14, C74-40, C76-06. C77-16, C79-18\}.

Due to time constraints, start-up companies choose people that fit the team by 
"character rather than skill" \{C63-6, 
C73-34, 
C76-6\}, implying that one's commitment and teamwork skills are more important than technical skills.

The reports discuss emotional issues of working in a start-up. Taking 
responsibility of many tasks at once creates anxiety, leading to burnout and 
loss of motivation. The reports suggest that the founders loss of motivation to 
continue 
operating a start-up, leads also to the end of the company.
Anxiety and burnout are also reported as outcomes and poor work and personal life balance \{C01-18, C27-35, C73-8\}.

Software engineering is an inherently human and team based intellectual activity. Team factors have emerged as critical in many different development environments and have effect to nearly any other activity~\cite{Fagerholm2012,Khan2011,Chow2008b,Sudhakar2012,Unterkalmsteiner}. However, as shown in Fig.~\ref{new_overview}, the reports suggest that software design is the most affected software engineering knowledge area. Difficulties in communication, inadequacies in skills and decision making are exposed through sub-optimal software design and poor product quality. This finding is consistent with Giardino et al.~\cite{Unterkalmsteiner} reporting that team's disregard of structures and engineering processes lead to deterioration of product architecture.

Fagerholm et al.~\cite{Fagerholm2015} reports a study on different factors affecting developer performance in lean and agile environments. In this study performance is used to benchmark efficiency and effectiveness of a team. They explore different factors facilitating performance, creating performance awareness, disrupting performance and others. This study shows that there are few key factors contributing to good developer experience. For instance, control of own work, decision power, and good environmental atmosphere contributes positively to overall team performance. Several factors, such as open office, collaboration and competition, and subordinance can both enhance and worsen a team performance.

Melo et al.~\cite{DeMelo2013} explores agile team productivity and lists several team related processes contributing to productivity, staff turnover and commitment. For example, good conflict management, sharing of expertise, and team coordination are essential to high developer commitment, low staff turnover and high productivity.

As exemplified with these two studies, state-of-the-art identifies the key ingredients for high performing teams in agile and lean environments. The same factors could be applicable in start-up teams. However, start-ups face certain specific limitations in team formation. Firstly, start-ups are founder centric and team environment highly depends on dynamics between the founders~\cite{Criaco2014}. Secondly, lack of resources limit access to highly skilled individuals, especially in the early stages. This could lead to sub-optimal initial team composition, and more effort is required to develop the team as a whole to reach the desired performance level~\cite{Fagerholm2015,Unterkalmsteiner}.

\subsection{Software Quality knowledge area} \label{RQ2_quality}

Software quality is a multi-faceted concept in software engineering. Nearly all 
other knowledge areas aim to somehow contribute to software 
quality~\cite{Society}. Kitchenham et al.~\cite{Kitchenham1996} identifies five 
perspectives on quality: transcendental, user, manufacturing, product and 
value-based view. 

Fig.~\ref{all_ka_frequency} illustrates that little details are provided in the 
reports regarding software quality. Superficial statements indicate that 
start-up companies see 
software quality as product's characteristics to meet users needs, thus 
focusing on the user perspective of quality \cite{Kitchenham1996}. The reports 
discuss usability, especially performance, user experience and reliability as 
their focus areas \{C50-15, 
C50-22, 
C61-17, 
C67-23, 
C67-26, 
C67-32, 
C78-15, 
C86-13, 
C86-14, 
C87-22, 
C37-4, 
C46-10\}.

The reports reflect on issues stemming from product quality: poor product 
reception due to an insufficient level of quality, or emphasis on the wrong 
quality aspects (for example, scalability over time-to-market) \{C14-15, 
C63-54,  C66-35,  C54-35, C69-12 \}.

Poor product adoption or loss or reputation are reported as consequences of 
poor product quality. However, very little details are provided on quality 
requirements and on any quality assurance procedures.

\paragraph{Discussion:}
While start-ups discuss very little how to achieve software quality, 
the consequences of inadequate quality, such as ruining the 
product's image or overly expensive product maintenance, are discussed. We 
observed that software quality is more discussed in closed companies. This 
tendency suggests that the importance of software quality 
may be realized only in hindsight. However, we could not confirm any 
statistically significant connection between company outcome and statements 
pertaining software quality.

As shown in Fig.~\ref{new_overview}, quality originates from product design. 
The 
analysis of the reported software design practices in Section 
\ref{rq2_sw_design_ka} reveals that companies often put excessive resources on 
improving certain quality attributes with little market need. The analysis of 
the reported requirements engineering practices in Section \ref{sec:summary_RE} 
shows that there is little discussion on quality requirements. 
Moreover, no practices to assure quality requirements were identified (see 
Section~\ref{RQ2_testing_ka}).

These findings indicate a gap between requirements engineering and software 
design. We argue that vague and wrongly prioritized quality requirements 
contribute to inadequate product design affecting the product's potential to 
deliver the promised value.

Regnell et al. \cite{Regnell2008} argue that in a market-driven context, product quality aspects have different thresholds. If a quality indicator is below a 
certain threshold, a product is useless. Within certain thresholds the product 
is useful but does not differentiate itself from competition. Above a certain 
threshold the product becomes competitive and at some point the 
quality becomes excessive and costly \cite{Regnell2008}. The concept of quality 
thresholds enables a company to identify important quality indicators and to 
perform requirements elicitation to determine the 
threshold values. Understanding of the threshold values and multiple 
perspectives of value enables the company to set and specify quality goals. 
Azar et al. \cite{Azar2007} propose a lightweight method to balance multiple 
influences to quality requirements and to determine optimal product quality 
goals. As elaborated in Section~\ref{rq2_sw_design_ka}, companies often fail to 
determine the required level of quality and waste resources on excessive 
quality features. More research is required to understand to what extent 
value-oriented requirements engineering practices could be applied in start-ups.

\subsection{Software Engineering Management knowledge area}\label{sec_management_ka}

The engineering management knowledge area concerns organizational aspects of 
software engineering, such as initiating, planning, monitoring, controlling and 
reporting of software engineering activities~\cite{Society}. While the dynamic 
environment in start-ups makes any detailed plans outdated quickly, the 
engineering process still must be controlled, follow resource and time 
constraints, and produce a result that is aligned with overall goals of the 
company.

With respect to the software engineering management knowledge area, the reports 
discuss effort estimation, monitoring and product discontinuation practices.

The reports suggest that the companies aim to achieve certain business goals, 
either to qualify for further external funding or to establish sufficient 
cash-flow to support development efforts. Pursuing these goals require 
investments in product development. Thus, estimation of the required 
resources is an important step to assess feasibility of the goals. The reports 
discuss how overly optimistic estimates contribute to the collapse of a company 
due to the lack of resources to finish the product \{C29-35, 
C66-35, 
C77-5, 
C79-11, 
C79-120\} or missed market opportunities \{C29-35, 
C79-11\}. However, the reports do not discuss any specific effort estimation method.

The start-ups report that any initial plans are based on assumptions \{C01-77, C02-74, C06-8,C52-30,1967-1978,C24-12\} and are adjusted as data from requirements validation comes in 
\{
C39-9,
C06-6,
C82-7,
C50-8,
C57-18,
C34-5,
C65-21,
C65-14,
C35-22,
C14-47,
C66-20,
C37-12,
C61-10,
C61-21,
C61-17,
C29-40,
C33-27,
C04-92,
C53-110,
C54-33,
C02-75,
C02-91\}.
Feedback from customers helps to determine the next immediate step, e.g. to 
improve certain features or collect feedback from a different stakeholder 
group. Even though the companies frequently refer to adjusting their plans 
based on success of a product release, we found very little details about this 
process \{C01-43, C12-142, C22-04, C02-32, C29-07, C33-44, C52-08, C52-20, 
C53-50, C46-13, C60-40, C82-16, C86-10, C51-12\}.

The reports suggest that start-ups attempt to estimate their progress by 
looking at various metrics that should be carefully selected and tied to 
business goals. Measuring the wrong thing or measuring too many things may lead 
to data overload and 
difficulties to interpret many conflicting measurements \{C01-66, 
C02-93, 
C05-5, 
C14-13, 
C29-25, 
C33-46, 
C35-26, 
C48-5, 
C50-18, 
C57-12, 
C64-17, 
C74-36\}. As Company \#50 reflects on selecting KPI: \myquote{``As a business 
leader you need to figure out the metric that matters most for your company and 
understand that the more you measure, the less prioritized you'll be. Don't 
fall into the trap of trying to measure everything.''}

The reports discuss different approaches to product discontinuation. Companies 
with 
products that had attracted a significant number of users report a timely 
notification to the users about the product discontinuation, and instructions 
on 
how to back-up their data and migrate to different solutions 
\{C18-4,C47-6\}. Some companies leave the product accessible but cease any further development 
or maintenance efforts \{C05-7, C10-15,C32-4,C36-5\}. Few companies report open-sourcing the product \{C80-13, C65-41\}.

The reports suggest that decisions stemming from engineering management 
influence release planning, e.g. by determining the release strategy, and 
requirements prioritization by setting prioritization goals. Engineering 
management also influences the professional practice knowledge area by setting 
expectations and constraints on the engineering team.

\paragraph{Discussion:}
Shahin \cite{Shahin2007} proposes criteria for defining organizational goals 
and a structured method to select key performance indicators for assessing 
progress towards said goals. A combination of SMART (Specific, Measurable, 
Attainable, Realistic and Timely) criteria to define goals and analytical 
hierarchy process (AHP) to select metrics to assess progress towards the goals 
\cite{Shahin2007} is a feasible alternative to the ``gut feeling'' approach 
described by Olsson et al. \cite{Olsson2015} and Terho et al. \cite{Terho2015}.

Terho et al. \cite{Terho2015} argue that fundamental changes to the start-ups' 
business plans, i.e. pivoting, are largely based on a gut feeling and are 
caused by an urgent need, e.g. need for more revenue. However, the collection 
on 
operational data (key performance indicators) helps to make more motivated 
decisions with specific goals \cite{Terho2015}. Therefore, use of a structured 
method to select important metrics, for example Shahin et al. 
\cite{Shahin2007}, could improve decision making in start-ups.

Garengo et al. \cite{Garengo2005} report that the lack of resources, little 
attention to formalization and a reactive approach are factors that hinder 
implementation of performance indicators in small and medium enterprises. Cocca 
et al. \cite{Cocca2009} argue that key performance indicators are essential to 
make informed decisions and propose best practices in implementation of 
performance indicators. The study lists qualities of a good performance indicator and exemplifies maturity grids as a tool to in decision making.
Shanin et al. 
\cite{Shahin2007} propose a lightweight technique based on the analytical 
hierarchy process (AHP) to select and prioritize key performance indicators. 
These practices for selecting and implementing key performance indicators could be considered for adaptation in start-ups \cite{Shahin2007,Cocca2009}.

Giardino et al. \cite{Giardino2014} emphasize the uncertainty in 
the start-up environment and argue that development teams in start-ups are 
formed by low-experienced engineers. The lack of joint and individual 
experience makes the application of expert judgement based effort estimation 
methods difficult \cite{Molokken2003}. Usman et al. \cite{Usman} report that 
the most widespread effort estimation technique in agile teams is planning 
poker, and the 
most popular size metric is story points. Moreover, the most common planning 
levels are current iteration and release. Whether the same methods are equally 
widespread in start-ups as well requires further research. However, group 
estimates, i.e. planning poker, is reported as being more accurate than 
individual estimates and could be very well applied in start-ups 
\cite{Haugen2006}.

Existing literature presents very little discussion on software product 
discontinuation. Jansen et al. \cite{Jansen2011} presents a structured plan on 
how to discontinue a software product. The proposed plan includes adequate 
pre-planning, transferring customers and partners to another solution and 
finally reallocation of the product team. Given the lack of resources it is 
unlikely that start-ups put more than absolute minimal effort on product 
discontinuation. However, the list of steps to discontinue a product proposed 
by Jansen et al. \cite{Jansen2011} could serve as a roadmap for product 
discontinuation in start-ups.

\subsection{Software Testing knowledge area}\label{RQ2_testing_ka}

Software testing is the dynamic verification that a product works as expected 
on a set of selected test cases. Some of the main tasks of testing are to 
determine what to test, specify input data and expected software behaviors, and 
to organize the process of software testing~\cite{Society}.

The reports contain only general statements addressing the software testing 
knowledge area. The statements suggest that 
start-up companies perform testing activities only when obvious issues emerge. 
For example, when performance had degraded below an acceptable level 
\{C24-16, 
C67-26, 
C86-13\}. Feedback from users is used to spot discrepancies in the product 
instead of performing rigorous internal testing \{C14-23, 
C29-36, 
C67-26\}.

Some companies report product failures in operation with substantial loss of resources and 
reputation. As company \#54 states: \myquote{``Finally, the server went down, 
scuttling the entire operation. Hill started handing out margaritas by the 
fistful to keep everyone happy. [..] The app picked up a number of 1-star 
reviews following the debacle''}. The reports did not specify whether the 
failures were due to lack of specific requirements or failure to meet such 
requirements.

As illustrated in Fig.~\ref{new_overview} software testing has a direct impact 
on software quality. The product must have an acceptable level of quality on 
all relevant aspects, or the product is simply useless to 
customers~\cite{Regnell2008}.

\paragraph{Discussion:}
Giardino et al. \cite{Unterkalmsteiner} argue that product 
quality has a low priority in software start-ups. 
Instead of rigorous internal testing, start-up companies utilize user feedback 
to determine if a level of quality is acceptable. A possible explanation is 
that due to frequently changing or unclear requirements there is no other 
reliable input for testing \cite{Graham2002}. However, as elaborated in Section 
\ref{RQ2_quality}, inadequacies in product quality can severely damage a 
product's reputation. Therefore, a company must 
carefully assess the risks stemming from the reliance on user side testing.

Another possible explanation is that a large part of testing is done by developers during the development 
process. This explains why the product appears to be in shape when released 
(code defects are removed), however failures in operation indicate a lack of 
design and stress testing \cite{Runeson2006}. Our findings suggest that start-ups are overlooking a potentially important knowledge area.

\subsection{Software Maintenance knowledge area}\label{RQ2_maintenance_ka}

A result of software development is a delivery of a software product to is users. However, post delivery defects may emerge, operating environment change or users could propose new requirements. The software maintenance phase begins when software is released to customers and ensures that the software continue to operate as intended. Software maintenance activities fall into perfective maintenance (to improve some quality the software, e.g. performance), corrective maintenance (to remove defects), adaptive maintenance (to adapt the software to a changed environment) and preventive maintenance (to prevent problems before they occur)~\cite{Society}.

Start-up companies report on software maintenance costs \{C23-3, 
C32-3\} and resource allocation for maintenance activities. The reports discuss the struggle of performing 
timely corrective maintenance due to understaffed teams. Long response time to 
product faults is reported as having a negative impact on product adoption 
\{C77-3\}. Adaptive maintenance to keep up with the product and any 
third-party component changes is reported as a concern \{C14-12, 
C38-9\}.

\paragraph{Discussion:}
The reports contain very little details on how start-ups manage software 
maintenance. However, inadequately high costs of keeping the product running is 
reported in relation to poor software design. As discussed in 
Section~\ref{rq2_sw_design_ka}, goals of software design shift from faster 
time-to-market to reducing maintenance effort. This shift takes place when the 
product feature set stabilizes and more and more users start-using the product 
\cite{Crowne2002}.

If this shift is not executed properly, a large number of users can overwhelm 
the product, exposing any inadequacies in product design and quality. Tackling 
these inadequacies may require substantial resources and time, contributing to 
the collapse of the company.

Webster et al. \cite{BatistaWebster2005} propose a taxonomy for evaluating 
risks pertaining to software maintenance. The taxonomy could be used in 
start-ups to identify and address potential maintenance risks during product 
development. Tom et al. \cite{Tom2013} argue that taking risks in engineering, 
i.e. creating technical debt, is a trade-off between shorter time-to-market and 
internal product quality. However, how exactly technical debt is handled in 
start-ups and to what extent this taxonomy is exhaustive and relevant in 
start-up context requires more research \cite{BatistaWebster2005,Tom2013}.

\section{Conclusions and future work} \label{sec_conclusions}

This study is the largest (by a number of studied cases) and broadest (by addressed software engineering knowledge areas) investigation into engineering aspects of start-ups to-date. With this study we paint a rich picture on how start-ups reflect on utilizing software engineering, what engineering practices start-ups use, and why. This study is aimed to characterize software engineering in start-ups, thus providing the necessary groundwork for conducting further and more detailed investigation into software-intensive product engineering in start-up context. To achieve our goal we perform third level analysis of start-up experience reports from 25 relatively successful and 63 closed start-ups.

Our results show that start-ups apply market-driven requirements engineering practices to discover and validate ideas for innovative products. However, the applied requirements engineering practices are often rudimentary and lack alignment with other knowledge areas. As a consequence, inadequacies in requirements engineering hinder other engineering activities and might lead to unwanted technical debt, poor product quality, and wasted resources on building irrelevant features. Further work is needed to identify good requirements engineering practices in start-ups.

We have found very little discussion regarding software testing. However, the reports discuss disastrous events when a product had failed in hands of customers. We conclude that software testing practices could be overlooked by start-ups. Further research is needed to understand state-of-practice in software testing in start-up context.

Other software engineering knowledge areas have a supportive role in continuous requirements identification and validation. For example, software design knowledge must support fast evolution of product prototypes, used to gather customer requirements, to a robust solution for easy maintenance.

The results of this study are intended to be useful to researchers in 
supporting further research in the area. The results can also be useful to 
start-up engineers willing to learn from experience of others. We have analyzed 
our findings in context of related work, thus hinting practitioners towards 
potentially useful practices. Future work includes examining key knowledge 
areas in more detail, and exploring to what extent the use of certain practices 
contributes to achieving start-up goals.

\section*{Acknowledgments}
The authors would like to thank Dr. Krzysztof Wnuk for insightful discussions, comments and hints to related work.

\bibliographystyle{spmpsci}      


\end{document}